\documentstyle[aps,prb,epsf]{revtex}
\begin{document}
\draft
\title { Hole motion in an arbitrary spin background:     
         Beyond the minimal spin-polaron approximation }
\author{ Matthias Vojta and Klaus W. Becker }
\address{Institut f\"{u}r Theoretische Physik,
         Technische Universit\"{a}t Dresden, D-01062 Dresden, \\ Germany}
\maketitle

\begin{abstract}
The motion of a single hole in an arbitrary magnetic background 
is investigated for the 2D $t$-$J$ model.
The wavefunction of the hole is described within a generalized 
string picture which leads to a modified concept
of spin polarons.
We calculate the one-hole spectral function using a large string basis
for the limits of a N\'{e}el ordered and a completely disordered 
background. In addition we use a simple approximation to interpolate
between these cases.
For the antiferromagnetic background we reproduce the well-known 
quasiparticle band. In the disordered case the shape of the spectral function
is found to be strongly momentum-dependent, the quasiparticle weight
vanishes for all hole momenta.
Finally, we discuss the relevance of results for the lowest energy eigenvalue 
and its dispersion obtained from calculations using a polaron of minimal size
as found in the literature.
\\
\end{abstract}


\section{Introduction}
Since the discovery of the high-temperature superconductivity the hole 
motion in strongly correlated electronic systems has attracted much interest.
It is widely accepted that many properties of the superconducting
cuprates are determined by the hole-doped CuO$_2$ planes.
Angle-resolved photoemission (ARPES) experiments
at zero\cite{Wells95} or small doping\cite{Dessau93,Marshall96}
indicate a quasiparticle band with a small bandwidth providing 
evidence for strong electronic correlations in the 
high-$T_c$ compounds.

The undoped materials are known to be antiferromagnetic Mott--Hubbard 
insulators. Neutron scattering experiments
show that their 
magnetic behavior can be well described by a spin $S={1 \over 2}$ 
Heisenberg model on a square lattice. 
The doped materials exhibit a strong dependence
of magnetic properties on the hole concentration $\delta$
in the $\rm CuO_2$ planes. With increasing hole concentration both
the N\'{e}el temperature and the staggered magnetization
decrease and vanish at a critical hole
concentration $\delta_c$ of a few percent before the system
becomes paramagnetic and metallic (or superconducting at sufficiently low
temperatures). 
Antiferromagnetic correlations are present also beyond the magnetic 
phase transition, e.g., in La$_{2-x}$Sr$_x$CuO$_4$ at $\delta=4\%$ 
($\delta_c \approx 2\%$) the magnetic correlation length is 
about 20 {\AA}.

The magnetic long-range order is already destroyed at a rather
small hole concentration $\delta$. Therefore exists a parameter
regime where a small number of holes move in a spin background which
is only short-range ordered. 
Here hole-hole correlations play no dominant role.
This motivates the investigation of
a single hole moving in an arbitrary spin background.
It is assumed that the holes move independently, i.e., 
that there is no phase separation or hole-binding.

The motion of a hole in a paramagnetic background is also relevant for
the study of the copper-oxide compound Ba$_2$Cu$_3$O$_4$Cl$_3$.
Its Cu$_3$O$_4$ planes consist of two copper subsystems with different
N\'{e}el temperatures \cite{Golden}. 
Therefore one finds a temperature range where
one subsystem exhibits antiferromagnetic order whereas the other one shows
paramagnetic behavior.

The low-energetic degrees of freedom of the copper-oxide planes
are believed to be well described by
the two-dimensional $t$-$J$ model\cite{Anderson87,Zhang88}:
\begin{equation}
H\, =\, - t \sum_{\langle ij\rangle \sigma}
      (\hat c^\dagger_{i\sigma} \hat c_{j\sigma} +
       \hat c^\dagger_{j\sigma} \hat c_{i\sigma})
    + J \sum_{\langle ij\rangle} \ ({\bf S}_i {\bf S}_j - {{n_i n_j} \over 4} )
\,.
\label{H_TJ}
\end{equation}
Here, ${\bf S}_i$ is the electronic spin operator and $n_i$ the electron
number operator at site $i$. The symbol $\langle ij \rangle$ refers to
a summation over pairs of nearest neighbors.
At half-filling the $t$-$J$ Hamiltonian reduces to the antiferromagnetic
Heisenberg model.
The electronic creation operators $\hat c_{i \sigma}^{\dagger}$
exclude double occupancies:
\begin{equation}
\hat c^{\dagger}_{i\sigma} = c^{\dagger}_{i\sigma} (1-n_{i,-\sigma})
\,.
\end{equation}

The hole motion within the $t-J$ and related models has been subject
of numerous investigations, see e.g. review articles\cite{Dagotto94,Brenig95}.
Special attention has been payed to the undoped case, i.e., to the case
where the spin background is given by the ground-state of the 2D Heisenberg
antiferromagnet. 
The main contribution to the hole motion in such an
antiferromagnetic background can be understood as follows: 
the hopping hole locally destroys the antiferromagnetic spin order 
leaving behind a string of spin defects. Quantum spin fluctuations
can repair pairs of frustrated spins. These processes
lead to a coherent motion of the hole in each of the two sublattices.
For $t/J > 1$ the bandwidth of this coherent hole motion is of order 
$J$ because the spin-flip part of the Hamiltonian (\ref{H_TJ}) is necessary
to remove the spin defects caused by hopping.

However, a spin background with more ground-state fluctuations allows for 
additional hole motion processes with an energy scale $t$. With increasing
spin disorder one expects the hopping term to become the dominant
contribution to the hole motion. 
The special cases we consider are
the completely disordered or random background (R) with 
$\langle {\bf S}_i \cdot {\bf S}_j \rangle = \frac {3}{4} \delta_{ij}$
and the ferromagnetic background (F)
$\langle {\bf S}_i \cdot {\bf S}_j \rangle = \frac{1}{4} + \frac {1}{2} \delta_{ij}$.
Note that the hole motion in a random background is a non-trivial situation
even in the case $J\rightarrow 0$ since we are still dealing with strongly
correlated electrons, i.e., with the restriction of no doubly occupied
sites.
The ferromagnetic background is easily discussed: 
A hole (with appropriate spin) can hop like a free particle through the
lattice without changing the ferromagnetic background. 
The only difference to a free electron is a negative phase for the hopping
term, i.e. $t \rightarrow -t$, since the object is a hole rather than
a particle.
The bandwidth for the hole motion is $8t$ (in two dimensions).

Up to now, the hole motion in weakly ordered or disordered spin backgrounds
has been studied analytically only in a few papers
\cite{BrinkRice70,MetzSchVoll92,HallMuBa95,HaRiYu95,BelRi95}.
Brinkman and Rice\cite{BrinkRice70} used the retracable path approximation
to discuss N\'{e}el-type (AF), random (R) and ferromagnetic (F) backgrounds. 
This approach only includes hole walks being completely 
background-restoring which is only a good approximation for systems
with strong antiferromagnetic order. 
Metzner, Schmit, and Vollhardt\cite{MetzSchVoll92} employed a different
method to discuss these three types of magnetic backgrounds within the
Hubbard model for $U\rightarrow\infty$. 
They considered a hypercubic lattice in $d$ dimensions and 
determined the one-hole Green's function
exactly in the limit $d\rightarrow\infty$.
These calculations were extended in ref. \cite{HallMuBa95} for generalized
spin backgrounds in infinite dimensions.

One possibility for the analytical treatment of the one-hole problem in 
an arbitrary background is a variational ansatz for the hole wavefunction
using strings\cite{BrinkRice70,Nagaoka,Trugman88,ShrSig88}
of local spin deviations. 
The resulting object is a hole surrounded by a cloud of spin deviations.  
It is usually referred to as spin-bag quasiparticle or magnetic polaron.
This approach was reduced to a "polaron of minimal size" where 
the basis set includes only the bare hole and strings of length 1
\cite{HaRiYu95,BelRi95,HaYuLo93,HaBaSchuRi96}.
Usually the calculation is then restricted to the lowest energy eigenvalue 
only.
In this paper we use a generalized string picture to discuss the hole
motion in an arbitrary magnetic background. We go beyond the minimal 
polaron ansatz mentioned above and include strings with large
lengths in the wavefunction. Using Mori-Zwanzig projection technique
for these strings one can calculate the one-hole spectral function
providing much more information about the hole states than the 
lowest eigenvalue alone. We have evaluated the arising expectation 
values for some special cases. Based on these results we discuss
the quality of the minimal polaron approximation.

Besides these analytical approaches 
numerical studies have been carried out for the Hubbard
and $t-J$ models at larger hole concentrations or finite temperatures.
There one expects a spin background with larger fluctuations
compared to the undoped system. However, the system sizes
accessible to numerical methods leave many problems unresolved.
Furthermore, the consideration of a given arbitrary magnetic 
background is difficult to realize within exact diagonalization 
or Lanczos methods.

The paper is organized as follows: 
In Sec. II we briefly sketch the variational ansatz for the 
one-hole wavefunction in an arbitrary spin background. 
We show how to calculate the one-hole spectral function
using projection technique and discuss approximations used within the 
polaron approach.
In Sec.~III we consider two cases which can be treated exactly within the 
polaron ansatz: A N\'{e}el ordered background for arbitrary
$t/J$ and a completely disordered background for $J=0$.
Based on these results we present in Sec.~IV a simple interpolation between the cases of 
an antiferromagnetic, a random and a ferromagnetic background. 
The discussion of the lowest energy eigenvalue within the minimal 
polaron approximation is subject of Sec.~V.
We evaluate the dispersion of the lowest energy eigenvalue for 
different maximum polaron sizes and discuss the spectral weight
of the lowest energy eigenvalue depending on the spin background.
A short conclusion will close the paper.


\section{ Spin polaron approach and one-hole spectral function }

In this section we outline the idea of generalized spin polarons.
We consider the ground state $|\Phi\rangle$ of
a two-dimensional half-filled system of $S= {1 \over 2}$ fermions
where double occupancies of sites are forbidden.
We assume that the state $|\Phi\rangle$ has a given magnetic configuration
and is an eigenstate of $S_z^{tot}$.
This spin configuration can be either long-range ordered (AF or F), 
short-range ordered or completely disordered (R). 
The configuration depends on the model parameters, 
the hole concentration, possible additional frustrating interactions
and finite temperature, see e.g. refs\cite{HaRiYu95,HaBaSchuRi96}. 
In this paper we are not going to investigate the magnetic properties 
of this state depending on this system parameters, we rather prefer to
take the magnetic configuration as given.

Next we inject one hole (with fixed momentum) into the state $|\Phi\rangle$ 
and study its motion. 
Hopping processes will disturb the spin background in
the vicinity of the hole. These spin deviations can be described by generalized
string operators $A_{n,\xi}(i)$ which contain multiple hole hopping.
$n$ denotes the path length, $\xi$ is the individual shape of the
path and $i$ is the initial lattice site; 
$A_{n,\xi}(i)$ operating on a state with one hole at site $i$ moves the
hole $n$ steps away by shifting the spins along the path with the 
shape $\xi$ by one lattice spacing.
By $m_n$ we denote the number of paths with length $n$. For $n=1$ there 
are $m_1=4$ different path shapes, for $n=2$ there are $m_2=12$ paths
and so on.
Explicitly, the operators $A_{n,\xi}(i)$ are defined by:
\begin{eqnarray}
A_{0,1}\,\, &=& \,\, 1 \,,\nonumber\\
A_{1,\xi}(i)\,\, &=&\,\, \sum_{j\sigma}\hat{c}_{j\sigma}^{\phantom\dagger}
        \hat{c}_{i\sigma}^\dagger\, R_{ji}^\xi\, ,\nonumber\\
A_{2,\xi}(i)\,\, &=&\,\, \sum_{jl\sigma\sigma'}
        \hat{c}_{l\sigma'}^{\phantom\dagger} \hat{c}_{j\sigma'}^\dagger
        \hat{c}_{j\sigma} ^{\phantom\dagger} \hat{c}_{i\sigma} ^\dagger
        \, R_{lji}^\xi\, , \label{PATH_DEF} \\
A_{3,\xi}(i)\,\, &=&\,\, \sum_{jlm\sigma\sigma'\sigma''}
        \hat{c}_{m\sigma''}^{\phantom\dagger} \hat{c}_{l\sigma''}^\dagger
        \hat{c}_{l\sigma'} ^{\phantom\dagger} \hat{c}_{j\sigma'}^\dagger
        \hat{c}_{j\sigma}  ^{\phantom\dagger} \hat{c}_{i\sigma} ^\dagger
        \, R_{mlji}^\xi \, ,\nonumber\\
\ldots\quad&\quad&\nonumber
\end{eqnarray}
The matrices $R_{i_n ... i_1}^\xi$ allow the hole to jump along a path
of shape $\xi$:
\begin{eqnarray}
R_{i_n ... i_1}^\xi\,\, &=&\,\, \left\{
   \begin{array}{rl}
     1&\quad {\rm if\,} i_1, ..., i_n\,\, {\rm are\,connected\,by\,path\,of\,shape}\,\xi\\
     0&\quad {\rm otherwise}\end{array}\right.\quad .
\end{eqnarray}
Note that we include only self-avoiding paths.
Therefore so-called Trugman paths and other loops are not included
in (\ref{PATH_DEF}). They only arise in expectations values like 
$\langle\Phi| {\hat c}_{{\bf k}\sigma}^\dagger A_{m,\nu}^\dagger \, A_{n,\xi}
  {\hat c}_{{\bf k}\sigma} |\Phi\rangle$ and
$\langle\Phi| {\hat c}_{{\bf k}\sigma}^\dagger A_{m,\nu}^\dagger \,H\, A_{n,\xi}
  {\hat c}_{{\bf k}\sigma} |\Phi\rangle$.
Fig. 1 shows the first path shapes for n=0,1,2.

If the background state is a N\'{e}el state then the application of a
path operator $A_{n,\xi}$ leads to a string of $n$ overturned 
(mismatched) spins attached to the hole. 
In a general spin background we generate a
string of spin deviations since all spins along the path are shifted
by one lattice spacing. 
For the limiting case of a ferromagnetic background the application of 
a path operator moves the hole without changing the background.

\subsection{ Projection technique }

To investigate the hole motion we consider a Green's function for a single 
hole describing the creation of a hole with momentum $\bf k$:
\begin{equation}
G({\bf k},\omega) \;=\; \sum_\sigma
  \langle\Phi| {\hat c}_{{\bf k}\sigma}^\dagger {1 \over {z-L}}
               {\hat c}_{{\bf k}\sigma}
  |\Phi\rangle
\label{HOLE_GF}
\end{equation}
where $z$ is the compex frequency variable, $z=\omega+i\eta$, 
$\eta\rightarrow 0$. 
The quantity $L$ denotes the Liouville operator defined by $LA = [H,A]_-$ for arbitrary
operators $A$.
The correlation function (\ref{HOLE_GF}) can be evaluated using 
Mori-Zwanzig projection technique\cite{Mori,Zwanzig}. 
In the following we briefly sketch this method.
One considers a set of operators $\{B_\nu\}$ (which here contains the 
hole creation operator ${\hat c}_{\bf k}$) and defines
dynamical correlation functions
\begin{equation}
G_{\nu\mu}(z) \,=\, \langle\Phi|\, B_{\nu}^\dagger {1 \over {z-L}}
                     B_{\mu}^{\phantom\dagger}\, |\Phi\rangle
\,.
\label{ARBKORR}
\end{equation}
Using projection technique one can derive the following set 
of equations of motion for the correlation
functions $G_{\nu\mu}(z)$:
\begin{equation}
\sum_{\nu} \left( z\delta_{\eta\nu}-\omega_{\eta\nu}-\Sigma_{\eta\nu}(z) \right)
  \,G_{\nu\mu}(z)\,\,=\,\,\chi_{\eta\mu} \, .
\label{PROJ_GLSYS}
\end{equation}
Here, $\chi_{\eta\nu}$, $\omega_{\eta\nu}$, and $\Sigma_{\eta\nu}$ are 
so-called static
correlation functions, frequency terms, and self-energies, respectively.
They are given by the following expressions:
\begin{eqnarray}
\chi_{\eta\nu} \;&=&\;
  \langle\Phi| \,B_{\eta}^\dagger B_{\nu}^{\phantom\dagger}\,
             | \Phi\rangle \, ,
  \nonumber\\
\omega_{\eta\nu} \;&=&\;
  \sum_{\lambda}  \langle\Phi|\,B_{\eta}^\dagger(LB_{\lambda}^{\phantom\dagger})\,
                  \, |\Phi\rangle \,\chi_{\lambda\nu}^{-1} \, ,
  \nonumber\\
\Sigma_{\eta\nu}(z) \,&=&\,\,
  \sum_{\lambda}   \langle\Phi|\,B_{\eta}^\dagger
                   \left( LQ {1 \over {z-QLQ}} QL B_{\lambda}^{\phantom +} \right)
                   |\Phi\rangle \,\chi_{\lambda\nu}^{-1} \, ,
\label{MATRIX_KUMDEF}
\end{eqnarray}
$\chi_{\nu\mu}^{-1}$ is the inverse matrix of $\chi_{\nu\mu}$, 
and $Q$ is defined by
\begin{equation}
Q = 1-P\, ,\quad
P = \sum_{\nu\mu} |B_{\nu}^{\phantom\dagger}|\Phi\rangle\,\,
                   \chi_{\nu\mu}^{-1} \,\,
                   \langle\Phi| B_{\mu}^\dagger|\,.
\end{equation}
$P$ denotes a projection operator projecting onto the subspace of the Liouville
space spanned by the operators $\{B_{\nu}\}$,
whereas $Q$ projects onto the complementary subspace. 

Using as dynamical variables the path operators $A_{n,\xi}$ defined above 
multiplied by ${\hat c}_{\bf k}$, i.e.,
\begin{equation}
B_{n,\xi} = \sum_\sigma A_{n,\xi} {\hat c}_{{\bf k}\sigma} \,,
\label{FULLVARSET}
\end{equation}
the one-hole correlation function 
we are interested in is the diagonal correlation function $G_{\nu\nu}$
with $\nu = (0,1)$.
With these dynamical variables the static and frequency matrices are
explicitly given by
\begin{eqnarray}
\chi_{m\nu,n\xi} \;&=&\; \sum_\sigma \:
  \langle\Phi| {\hat c}_{{\bf k}\sigma}^\dagger 
  A_{m,\nu}^\dagger\, A_{n,\xi}\,
  {\hat c}_{{\bf k}\sigma} |\Phi\rangle \, , \nonumber\\
\omega_{m\nu,n\xi} \;&=&\; \sum_\sigma \:
  \langle\Phi| {\hat c}_{{\bf k}\sigma}^\dagger 
  A_{m,\nu}^\dagger\,(L\, A_{n,\xi}\,
  {\hat c}_{{\bf k}\sigma})\, |\Phi\rangle
\label{HOLEPROJ_MATEL}
\end{eqnarray}
Often one neglects the self-energy terms which describe processes 
outside the subspace spanned by the dynamical variables 
(which are here the path operators).
This can be done if the set of relevant variables is chosen sufficiently
large to cover the essential part of the dynamical behavior of the 
system.

The poles of the correlation function (\ref{HOLE_GF}) correspond
to eigenstates of the object which is usually called "spin polaron".
Using projection technique one obtains the energies and the spectral
weight distribution for these poles.
This means that all spectra calculated
here are discrete spectra because the self-energy terms within projection
technique are neglected. 
(In the figures presented below we have introduced an artificial Lorentzian
broadening to plot the spectral functions.)
Therefore the present approach cannot account for
analytical features of the spectral functions as e.g. the nature of the
spectra at the band edges.

When calculating expectation values containing path operators $A_{n,\xi}$
as $\chi_{m\nu,n\xi}$ and $\omega_{m\nu,n\xi}$
the effect of hopping processes can be rewritten in terms of spin 
operators. The lowest matrix elements are:
\begin{eqnarray}
\sum_\sigma \:
\langle\Phi|{\hat c}_{i\sigma}^\dagger\,A_{0,1  }\, 
{\hat c}_{i\sigma}|\Phi\rangle \;&=&\;
   1 \,, \nonumber\\
\sum_\sigma \:
\langle\Phi|{\hat c}_{j\sigma}^\dagger\,A_{1,\xi}\, 
{\hat c}_{i\sigma}|\Phi\rangle \;&=&\;
   2 \, \langle\Phi| 
   \left( {\bf S}_j \cdot {\bf S}_i\,+\, \frac {1}{4} \right )
   |\Phi\rangle  \,
   R_{ji}^{(\xi)} \,,  \label{PATH_TO_SPIN} \\
\sum_\sigma \:
\langle\Phi|{\hat c}_{k\sigma}^\dagger\,A_{2,\xi}\, 
{\hat c}_{i\sigma}|\Phi\rangle \;&=&\;
   \langle\Phi|
   \left (         - {\bf S}_k \cdot ({\bf S}_j \times {\bf S}_i)
                \,+\,{\bf S}_j \cdot {\bf S}_i
                \,+\,{\bf S}_k \cdot {\bf S}_i
                \,+\,{\bf S}_k \cdot {\bf S}_j
                \,+\, \frac {1}{4} 
   \right )             
   |\Phi\rangle \,
   R_{kji}^{(\xi)} \,,\nonumber\\ 
\quad &\,&...
\nonumber
\end{eqnarray}
In this way all matrix elements from (\ref{HOLEPROJ_MATEL}) 
can be transformed into expectation values of multi-products of
spin operators formed with the background state $|\Phi\rangle$. 
Note that this is the only point where the properties of the magnetic background 
enter:
For the investigation of the one-hole motion
one has to know static multi-point spin correlation functions. 
For a general spin background the knowledge of all 
these many-point correlators is usually not available. 
One way to evaluate these functions could be factorizing them into two-point 
(or two-point and four-point) correlation functions, see e.g. refs.
\cite{HaRiYu95,BelRi95}. For the special backgrounds discussed below
a factorization is not necessary.

Next we shall discuss some approximations which have to be done within the
polaron scheme. 
In most practical calculations one can take into account
only a finite number of path operators up to a certain length $n_{max}$.
This truncation should be possible if the weight of the path states 
decreases rapidly with increasing path length, i.e., 
if the polaron is localized in space and
longer paths do not contribute to the wavefunction, see also next subsection.
This is certainly true only for low-lying polaron states in an antiferromagnetic
background and for finite antiferromagnetic exchange $J>0$.
In this case the hopping hole creates frustrated spins in the antiferromagnetic
background leading to an Ising potential which increases with increasing 
path length \cite{ShrSig88,VojBeck96,EdBeckStep90,BEW}.
The question is whether this truncated polaron method gives reliable 
results also in different magnetic backgrounds. 
One purpose of the present paper is to illustrate that it does:
We have calculated the one-hole spectral function for the random
spin background and different maximum path lengths up to 256. With
increasing path length we have found only minor changes in the shape of
the spectral function, i.e., beyond a path length of about 30 the
problem is almost converged. Furthermore, our results for the
random background coincide with those from refs. 
\cite{MetzSchVoll92,HallMuBa95}
which have been obtained using a completely different analytical 
approach.

A second approximation within the polaron method concerns the individual
paths. Especially when considering longer paths the number of 
individual path shapes increases rapidly ($m_n\approx 3^n$).
Therefore we are usually forced to consider 
only one variable per path length which is the sum of
all individual paths of this length, i.e., the set of dynamical variables
reduces to
\begin{equation}
B_n = \sum_{\xi=1}^{m_n} \sum_\sigma \:
A_{n,\xi} c_{{\bf k}\sigma} \,.
\label{REDVARSET}
\end{equation}
Here, all different paths with equal length have been given the same
weight which is a variational restriction compared to (\ref{POLGS_ANS}). 
However, if the hole momentum does not favor a particular direction, i.e., if
${\bf k}=(0,0)$ or $(\pi,\pi)$, and if one neglects geometrical differences 
between the paths of same length then the weights of these paths 
are in fact equal for the lowest polaron eigenstate.
This results in an s-like ground state of the quasiparticle. 
For general momenta the symmetry of the ground state is only nearly 
s-like. This point will be discussed further in Sec. V.
Higher polaron states can have angular nodes in the coefficients,
these states are not covered by the approximation (\ref{REDVARSET}).
However, we have found that for the one-particle dynamics these states 
give only minor contributions to the spectrum.
Only the (nearly) s-like states contribute considerable spectral weight
to the one-hole spectral function
leading to $n_{max}$ poles with non-vanishing weight. The reason lies
in the symmetry of the Hamiltonian: None of the four lattice directions are
preferred, so hopping processes in all directions carry equal weight.
We summarize that the reduced set of variables (\ref{REDVARSET}) gives 
almost the same result as (\ref{FULLVARSET}), at least for
momenta $(0,0)$ and $(\pi,\pi)$.

\subsection{ Ansatz wavefunction }

An alternative approach to the polaron problem consists of an ansatz 
wavefunction using the generalized path operators defined 
above (\ref{PATH_DEF}).
A variational wavefunction for one hole with momentum $\bf k$ can be
constructed as linear combination of path states:
\begin{eqnarray}
|\psi\rangle \;&=&\;
  \sum_{n=0}^{n_{max}}\sum_{\xi=1}^{m_n} \sum_\sigma \:
  \left(\lambda_{n,\xi} A_{n,\xi}\right) \,
  {\hat c}_{{\bf k}\sigma} \, |\Phi\rangle \,,
\label{POLGS_ANS} \\
A_{n,\xi}\;&=&\;\sum_i A_{n,\xi}(i) \,.
\nonumber
\end{eqnarray}
Inserting this ansatz (\ref{POLGS_ANS}) into the Schr\"{o}dinger equation 
we obtain a generalized eigenvalue problem:
\begin{equation}
  \sum_{n\xi} \sum_\sigma \,
  \lambda_{n,\xi}
  \langle\Phi| {\hat c}_{{\bf k}\sigma}^\dagger A_{m,\nu}^\dagger\,H\, A_{n,\xi}
  {\hat c}_{{\bf k}\sigma} |\Phi\rangle
  \;=\;
  E\:
  \sum_{n\xi} \sum_\sigma\,
  \lambda_{n,\xi}
  \langle\Phi| {\hat c}_{{\bf k}\sigma}^\dagger A_{m,\nu}^\dagger A_{n,\xi}
  {\hat c}_{{\bf k}\sigma} |\Phi\rangle
\,.
\label{POL_SGL}
\end{equation}
The matrix elements in (\ref{HOLEPROJ_MATEL}) are the same expressions
as in the eigenvalue problem (\ref{POL_SGL}).
So the physics covered by both methods is the same:
All processes which can be described by local spin deviations near the
hole are taken into account.
The poles of the Green's function (\ref{HOLE_GF}) equal the 
energy eigenvalues of the variational problem (\ref{POL_SGL}) (besides an
energy shift). 
The solutions of (\ref{POL_SGL}) are the eigenstates of the spin polaron.
An important additional information obtained via projection technique 
is of course the spectral weight distribution among the poles of 
$G({\bf k},\omega)$.

The approximation of using only one variable per path length, i.e., the
reduced set of dynamical variables (\ref{REDVARSET}) as described above,
corresponds to a restricted variational ansatz for the wavefunction:
\begin{eqnarray}
|\psi^{red}\rangle \;&=&\;
  \sum_{n=0}^{n_{max}} \lambda_n \sum_{\xi=1}^{m_n} \sum_\sigma \,
  A_{n,\xi} \,
  {\hat c}_{{\bf k}\sigma} \, |\Phi\rangle \,. 
\label{POLGS_REDANS}
\end{eqnarray}
The energy eigenvalues which are obtained using this wavefunction
are the same as the poles of the one-hole Green's function
calculated with projection technique and the reduced set of operators
(\ref{REDVARSET}).


\section{ Disordered and N\'{e}el backgrounds }

In this section we consider two cases where the expectation values 
(\ref{HOLEPROJ_MATEL}) can be calculated for arbitrary long paths. 
One is the completely disordered or random background (R)
without exchange interaction ($J=0$) where the strong correlations
are present only in the exclusion of double occupancies. 
The other one is a N\'{e}el ordered background (AF) (for any value of $J$).
In the numerical calculations we have used the full set of variables 
(\ref{FULLVARSET}) with paths up to length $n_{max}=5$ and the reduced set
(\ref{REDVARSET}) up to $n_{max}=256$.

\subsection{ Disordered background }

For the random background the expectation value of all spin 
correlation functions between different sites vanish.
Therefore expectation values of path operators are easily evaluated.
One finds (compare (\ref{PATH_TO_SPIN}) ):
\begin{eqnarray}
\sum_\sigma \,
\langle\Phi| {\hat c}_{j\sigma}^\dagger A_{n,\xi} 
{\hat c}_{i\sigma} |\Phi\rangle \:=\: 
  \frac {1} {2^n} \;R_{j...i}^{(\xi)}
\,.
\label{PATHEV_R}
\end{eqnarray}
This can be understood as follows: The expectation value of a path operator
is non-zero only if the hopping process does not change the spin background,
i.e., if all spins along the path are parallel aligned.
In the random background the probability for each spin of pointing up or down
is $\frac{1}{2}$. Therefore the probability of two neighboring spins being
parallel aligned is $\frac{1}{2}$. For a path of length $n$ which includes
$n+1$ spins the probability of all spins being parallel is $\frac {1} {2^n}$.

For $J=0$ all matrix elements in (\ref{HOLEPROJ_MATEL}) reduce to
expectation values of path operators since the hopping term $H_t$ only
changes the length of a path by one. 
The remaining non-trivial terms to calculate are the phase factors
$e^{{\rm i}{\bf k}({\bf R}_j - {\bf R}_i)}$ arising from expectation
values like 
$\langle\Phi| {\hat c}_{{\bf k}\sigma}^\dagger A_{n,\xi} 
{\hat c}_{{\bf k}\sigma} |\Phi\rangle$.
In principle such a phase factor has to be calculated for every 
individual path. For long paths the number of terms is very large
($\approx 3^{n_{max}}$), so we have performed Monte-Carlo sampling
over $10^5$ different paths (which is sufficient for statistical
errors $< 1\%$).

Results for the spectral function for different momenta are displayed 
in Fig. 2. The shape of the spectral function is strongly momentum
dependent. The spectral weight of the lowest pole vanishes which indicates
the absence of a quasiparticle (QP) peak.
Considerable spectral weight near the band minimum is present only
at momenta near $(\pi,\pi)$. 
The energy difference between the maxima at $(0,0)$ and $(\pi,\pi)$ 
is about $6.4 t$ which can be considered as a "bandwidth".

In a random spin background the probability of large ferromagnetically aligned 
spin clusters is non-zero (but goes to zero with increasing cluster size
in the thermodynamic limit).
In such a cluster the hole moves like a free particle. This should cause
the density of states to have tailes with exponentially small weight 
extending to the edges of the free-particle band, see also refs. 
\cite{BrinkRice70,MetzSchVoll92}.
So we expect the one-hole spectral function e.g. at momentum $(\pi,\pi)$ 
to have a tail down to $\omega = -4t$ (being the energy of the free hole) 
with vanishing weight at $\omega = -4t$.
However, such analytical features cannot be extracted from the
results of the projection-technique calculations presented here, see discussion
in Sec. II A. 
The tailes at the band egdes visible e.g. in Fig. 2
arise only from the artificial broadening of the lines and do 
not have a physical meaning.

Note that the results presented in Fig. 2 agree well with the ones obtained 
within $d \rightarrow \infty$ approaches\cite{MetzSchVoll92,HallMuBa95}.
From this agreement 
we conclude that the processes included in our calculation, i.e., local spin
deviations caused by hole hopping, cover the essential part of the 
dynamics of the one-hole motion in a disordered background. 
This is a non-trivial fact since the hole motion processes are not as
localized in real space as in the case of a N\'{e}el background (AF) with
non-zero Ising interaction.
In the R case paths with all lengths contribute to the hole states,
but a truncation of the subspace of path operators (with $n_{max}$ 
being sufficiently large) causes no essential changes in the 
one-hole spectral function.
The influence of the maximum path length $n_{max}$ included in the 
calculation is illustrated in Fig. 3.
It can be seen that beyond a maximum path length of 30 there are practically
no changes in the shape of the spectral function.

\subsection{ N\'{e}el background }

The hole motion in a N\'{e}el ordered background using strings of
spin defects has been discussed in a number of papers, see e.g.
\cite{ShrSig88,VojBeck96,EdBeckStep90,BEW}.
Hole motion on a self-avoiding path in the AF background always
creates spin defects, so the expectation value of a single
path operator vanishes:
\begin{eqnarray}
\sum_\sigma \,
\langle\Phi| {\hat c}_{j\sigma}^\dagger A_{n,\xi} 
             {\hat c}_{i\sigma} |\Phi\rangle \:=\:
\delta_{n,0}\,\delta_{ij}
\,.
\label{PATHEV_AF}
\end{eqnarray}
Non-zero matrix elements arise only from paths returning to their origin,
and from spin-flip terms present in the Hamiltonian.
Important matrix elements are the Ising energies due to frustration
\begin{eqnarray}
\sum_\sigma \,
\langle\Phi| {\hat c}_{i\sigma}^\dagger A_{n,\xi}^\dagger (L_{Ising} A_{n,\xi}
             {\hat c}_{i\sigma}) |\Phi\rangle \:=\:
{J \over 2} \, (2n + 3 - \delta_{n,0})
\end{eqnarray}
and the spin-flip contribution
\begin{eqnarray}
\sum_\sigma \,
\langle\Phi| {\hat c}_{j\sigma}^\dagger A_{n+2,\xi'}^\dagger (L_\perp A_{n,\xi}
             {\hat c}_{i\sigma} ) |\Phi\rangle \:=\:
{J \over 2} \, \delta_{\xi+2,\xi'} \: 
                \delta_{{\bf R}_j-{\bf R}_i, {\bf\Delta}_2}
\,.
\end{eqnarray}
Here, the symbol $\delta_{\xi+2,\xi'}$ means that path $\xi'$ has to be 
obtained from path $\xi$ by two further hopping steps, and
${\bf\Delta}_2$ is a lattice vector consisting of two hops.
For details concerning the evaluation of the matrix elements we refer 
to former publications\cite{VojBeck96,BEW}.

The spectral function obtained with the reduced set of variables
(\ref{REDVARSET}) is showed in Fig. 4 for $J/t=0.4$
and for different momenta. We observe a non-zero weight of the lowest pole
for all momenta indicating a quasiparticle peak.
As mentioned the present approach cannot give analytical information about 
the spectra at the band edges. 
So we cannot answer exactly the question for the quasiparticle
weight. However, a non-zero QP weight in the quantum AF ordered state 
(as well as in the N\'{e}el state considered here)
is in agreement with most finite-size scaling studies of numerical 
calculations and with some analytical investigations of the $t$-$J$ model
\cite{QPWeight}. 
Note that this is in variance with an argument by Anderson \cite{Anderson90}.

Fig. 5 shows the one-hole spectrum for $J=0$. Here the hole is localized
since we have no $H_{\perp}$ terms destroying spin fluctuations caused
by hole hopping. 
(The variable set does not cover loops like trugman paths.)
Therefore the spectrum is momentum-independent. No quasiparticle
peak is present because the coherent motion of the hole is suppressed.


\section{ Interpolation between antiferromagnetic and ferromagnetic 
          background }

Up to now we have considered three background states: antiferromagnetic (AF), 
disordered (R) (and $J=0$) and ferromagnetic (F). For the cases AF and R
we have calculated the spectral function using arbitrarily long paths, the
ferromagnetic case F is trivial. 
To see how the crossover of the quasiparticle behavior between these 
cases occurs we try a simple interpolation between the results
(\ref{PATHEV_R}) and (\ref{PATHEV_AF}).
We consider the following ansatz:
\begin{eqnarray}
\sum_\sigma
\langle\Phi| {\hat c}_{j\sigma}^\dagger A_{n,\xi} 
             {\hat c}_{i\sigma} |\Phi\rangle \:=\: 
 p^n \;R_{j...i}^{(\xi)}
\,.
\label{INTERP_ANS}
\end{eqnarray}
As explained above, non-zero contributions to the expectation value of 
a path operator arise only from hopping process along paths of
parallel aligned spins.
Thus $p$ can be seen as the probability for two neighboring spins to be parallel
aligned. This interpolation covers the three limiting cases:
\begin{equation} 
p\:=\: \left\{
   \begin{array}{rl}
     1  &\quad {\rm ferromagnetic\, background\, (F)} \\
     \frac{1}{2} &\quad {\rm random\, background\, (R)} \\
     0  &\quad {\rm antiferromagnetic\, background\, (AF)}
   \end{array} \right .
\end{equation}
In this way parameter values $0<p<0.5$ can be interpreted as weak
antiferromagnetic order, whereas $0.5<p<1$ describe weak ferromagnetic
order. 

Of course this simple ansatz does not distinguish between short-range
and long-range order. Furthermore, it does not cover the ground state
of the Heisenberg antiferromagnet, i.e., a N\'{e}el state with
quantum fluctuations. There we have 
$\langle\Phi|{\bf S}_j\cdot{\bf S}_i|\Phi\rangle\approx-0.303$ with
$i$ and $j$ being nearest neighbor sites. This leads to 
$\sum_\sigma \langle\Phi| {\hat c}_{j\sigma}^\dagger A_{1,\xi} 
{\hat c}_{i\sigma} |\Phi\rangle \approx
-0.053 \,R_{ji}^{(\xi)}$, compare (\ref{PATH_TO_SPIN}).
Nevertheless, the ansatz (\ref{INTERP_ANS}) shows the main effect 
of the crossover between the antiferromagnetic and disordered 
spin backgrounds.

To include the exchange interaction $J$ we employ
a simple mean-field treatment. The following matrix elements are
taken into account:
\begin{eqnarray}
\sum_\sigma \,
\langle\Phi| {\hat c}_{i\sigma}^\dagger A_{m,\nu}^\dagger (L_{Ising} A_{n,\xi}
             {\hat c}_{i\sigma}) |\Phi\rangle \:=\:
J\,(p-\frac{1}{2})\,(p-1)  \, (n+m + 2) \;
\sum_\sigma \,
\langle\Phi| {\hat c}_{i\sigma}^\dagger A_{m,\nu}^\dagger A_{n,\xi}
             {\hat c}_{i\sigma}) |\Phi\rangle
\label{MEAN_J}
\end{eqnarray}
These terms represent the change of the Ising energy due to spin 
deviations coming from hopping processes. They interpolate between the three
limiting cases AF, R, and F: In the antiferromagnetic background (AF)
the Ising energy increases linearly with the path length provided here
by the term $J\,(n+m+2)$. 
For both the R and F cases the change of the Ising energy via hole hopping
vanishes:
For the R case the Ising energy does not change since the average Ising 
energy is zero whereas in the F background all spins remain parallel aligned. 
Furthermore, the dependence of the Ising energy on $p$ near $p=0$ is 
quadratic. These two factors $p$ follow from the average Ising energy
being linear in $p$ and the probability of creating a spin defect via
hole hopping being also linear in $p$.

Results for the spectral function at momentum $(0,0)$, $J/t=0.4$ 
and different values of $p$ are shown in Fig. 6. 
At the bottom of the spectrum
quasiparticle weight develops with increasing antiferromagnetic order.
However, these quasiparticle peak is dominant only for strong
antiferromagnetism. 
So it is obvious that the position of this quasiparticle peak is
not the only relevant quantity, especially when comparing with 
photoemission experiments done at underdoped high-$T_c$ 
superconductors. 
Fig. 7 shows the spectra for $J=0$. In this case we have no quasiparticle
peak even for the antiferromagnetic background since coherent motion of
the hole due to the transverse part of the magnetic exchange is 
absent.


\section{ Discussion of the lowest energy eigenvalue }

In this section we consider the lowest energy eigenvalue obtained within
the polaron ansatz for the wavefunction.
The lowest eigenvalue has been calculated before \cite{BelRi95,HaYuLo93} for a 
"polaron of minimal size". This ansatz includes only paths up to length 1
for an antiferromagnetic as well as for a disordered background.
Two questions have to be answered: 
1) Is the lowest energy eigenvalue relevant in the sense that it 
carries spectral weight? 
2) Does the reduced ansatz provide enough basis states to give
reliable results compared with the "full" ansatz, e.g., with paths
up to length 256?

The first question has already been addressed in the last section.
The spectral weight of the lowest energy eigenvalue decreases with
decreasing antiferromagnetic correlations in the spin background state.
It seems to be non-zero for all states with antiferromagnetic 
correlations in the case $J>0$, but to resolve this subtle question clearly 
more work is necessary. 
However, the spectrum is dominated already for a weak antiferromagnetic 
background (or a disordered one) by structures 
which are not located at the band minimum. 
These dominant structures should be visible in ARPES experiments. 
Thus the lowest energy eigenvalue is only relevant for a background with
strong AF correlations, i.e., at very small hole concentrations
and low temperatures. It is completely irrelevant for a disordered
background. 

The second question concerns the truncation of the polaron wavefunction.
For a N\'{e}el-ordered background the minimal polaron ansatz 
with paths up to length $n_{max}=1$ fails
completely because the dominant hole motion process caused by $H_{\perp}$ is
not covered by paths with maximum length 1.
But also a wavefunction with maximum path length of $n_{max}=2$ gives a 
bandwidth ($E(0,0)-E(\pi/2,\pi/2)$)
being a factor of 2 too small as compared with a full ansatz
(for values of $J/t=0.2-0.5$). 
Fig. 7 shows the quasiparticle dispersion of the hole motion in
a N\'{e}el background for different maximum path lengths (calculated
with the ansatz (\ref{POLGS_ANS}) ).

Considering an antiferromagnetic background state with more ground-state
spin fluctuations one expects that the results for the hole dispersion 
obtained using the minimal polaron ansatz\cite{BelRi95,HaYuLo93} become better
with increasing spin fluctuations since nearest-neighbor hopping processes 
of order $t$ become more important.
However, the bandwidths obtained in these calculations are still not reliable.
The reason lies in the small number of basis states which do not provide
sufficient degrees of freedom for the variational wavefunction.
For detailed investigations we refer to a forthcoming 
publication \cite{VojBeck97_3}.

For a disordered background one can also evaluate the dispersion of
the lowest energy eigenvalue although it does not carry spectral
weight. With $n_{max}=0$ (only the bare hole as trial state)
one obtains $E(0,0)-E(\pi,\pi)=4 t$ ($J=0$), using $n_{max}=1$ one
finds $E(0,0)-E(\pi,\pi)\approx 1.5 t$, $n_{max}=2$ leads to
$E(0,0)-E(\pi,\pi)\approx 0.6 t$, and the "converged" value within
our calculations is $|E(0,0)-E(\pi,\pi)| < 10^{-2}t$.


\section{Conclusion}

In this work we have studied the dynamics of a single hole moving in 
an arbitrary spin background within the framework of the two-dimensional
$t$-$J$ model.
The one-hole spectral function has been calculated using Mori-Zwanzig projection
technique for a large set of path operators. 
These operators describe local spin deviations around the hole which
lead to a generalized picture of spin-bag quasiparticles or spin polarons.

We have calculated the one-hole spectral function using three limiting
cases for the spin background: antiferromagnetic (AF), 
disordered (R) (and $J=0$) and ferromagnetic (F). 
The obtained results for the AF case shows the features well-known from
numerical and analytical investigations on the one-hole problem,
i.e. a quasiparticle-like peak with a dispersion which has its
minima at $(\pm\pi/2,\pm\pi/2)$ and a bandwidth of about $2.2 J$.
For the disordered background the spectral function is strongly 
momentum dependent but shows no quasiparticle peak. These results
coincide with the ones of a $d\rightarrow\infty$ approach for the
$U\rightarrow\infty$ Hubbard model\cite{MetzSchVoll92,HallMuBa95}.
A hole in a ferromagnetic background behaves like a free particle.
Using a simple interpolation we have studied the
crossover between the AF, R and F situations.
The present calculation neglects processes outside the subspace of path
operators. These processes would provide lifetime effects due to the
scattering of the polaron states with spin waves. However, we
expect no essential changes in the spectral function.

Finally, we have discussed the quality of the "minimal polaron
approximation" often used in the literature. This reduced description
consists of truncating the ansatz for the polaron wavefunction to
paths with maximum length 1 and calculating the lowest energy eigenvalue
only.
In most cases this approximation is not sufficient to cover essential 
properties of the hole motion process.

\acknowledgements

{It is pleasure for us to thank E. Dagotto and R. Hayn for helpful 
discussions.}


\newpage

\begin{figure}

\epsfxsize=15cm
\epsfysize=11cm
\epsffile{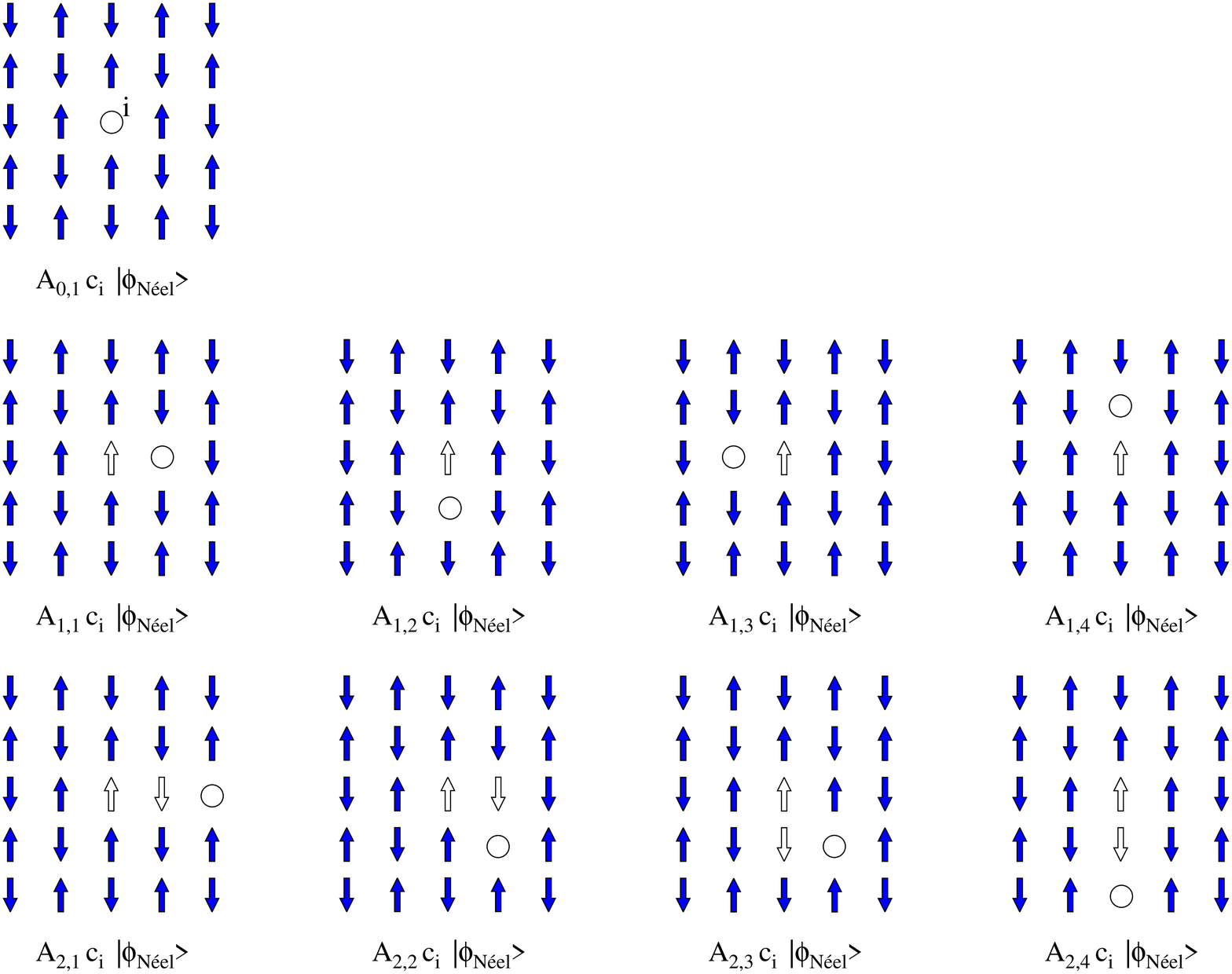}
\caption{ The first path shapes of lengths $n=$ 0,1, and 2 
          created by the operators $A_{n,\xi}$
          acting on a hole at site $i$. }
\vspace{1cm}

\epsfxsize=13cm
\epsfysize=8cm
\epsffile{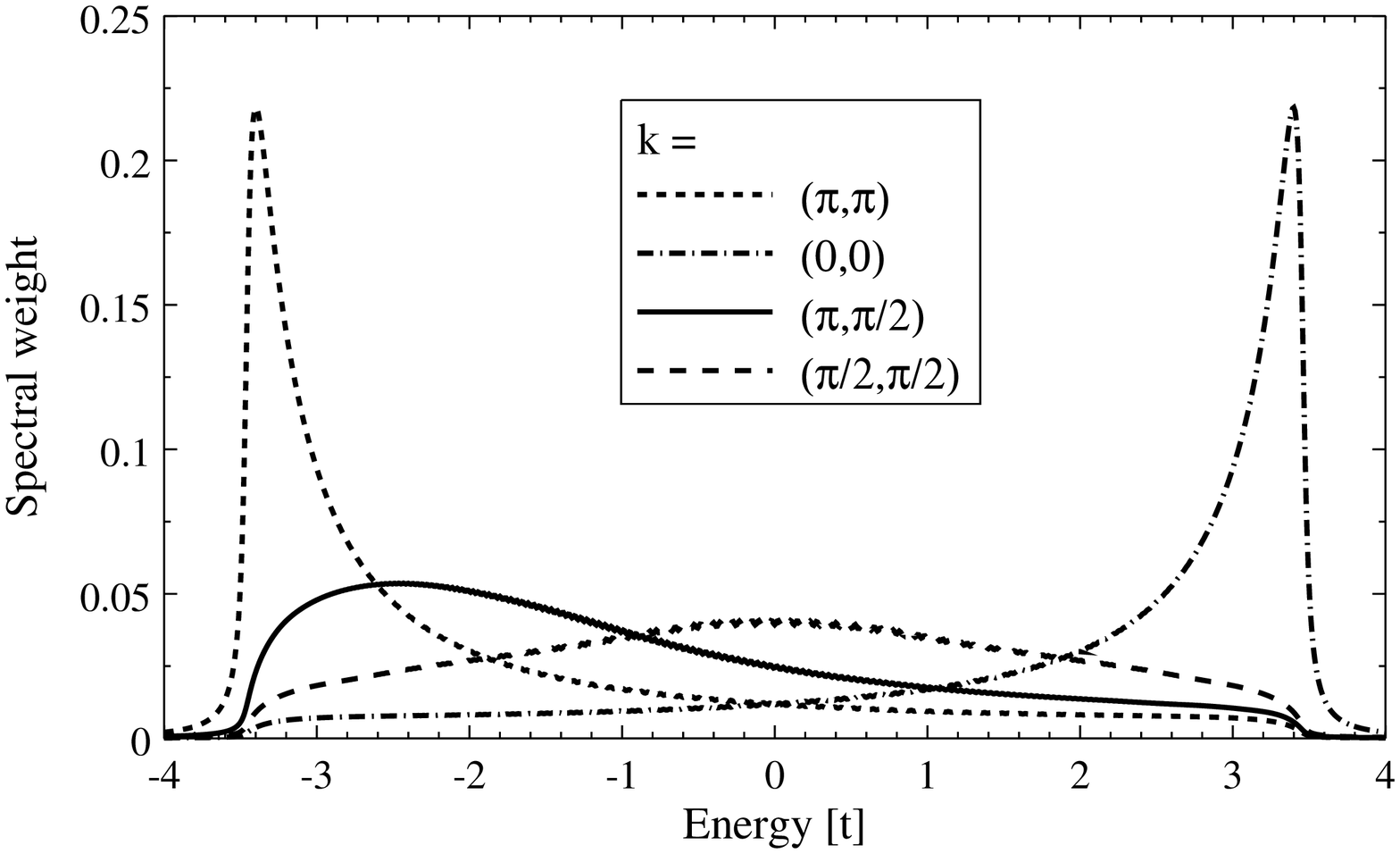}
\vspace{1cm}
\caption{ One-hole spectral function for the random spin background
          and $J=0$. Curves are shown for different hole momenta.
          The reduced variable set (\protect\ref{REDVARSET})
          including paths up to length 256 has been used.
          We have introduced an artificial broadening of $10^{-3} t$. }
\newpage

\epsfxsize=13cm
\epsfysize=8cm
\epsffile{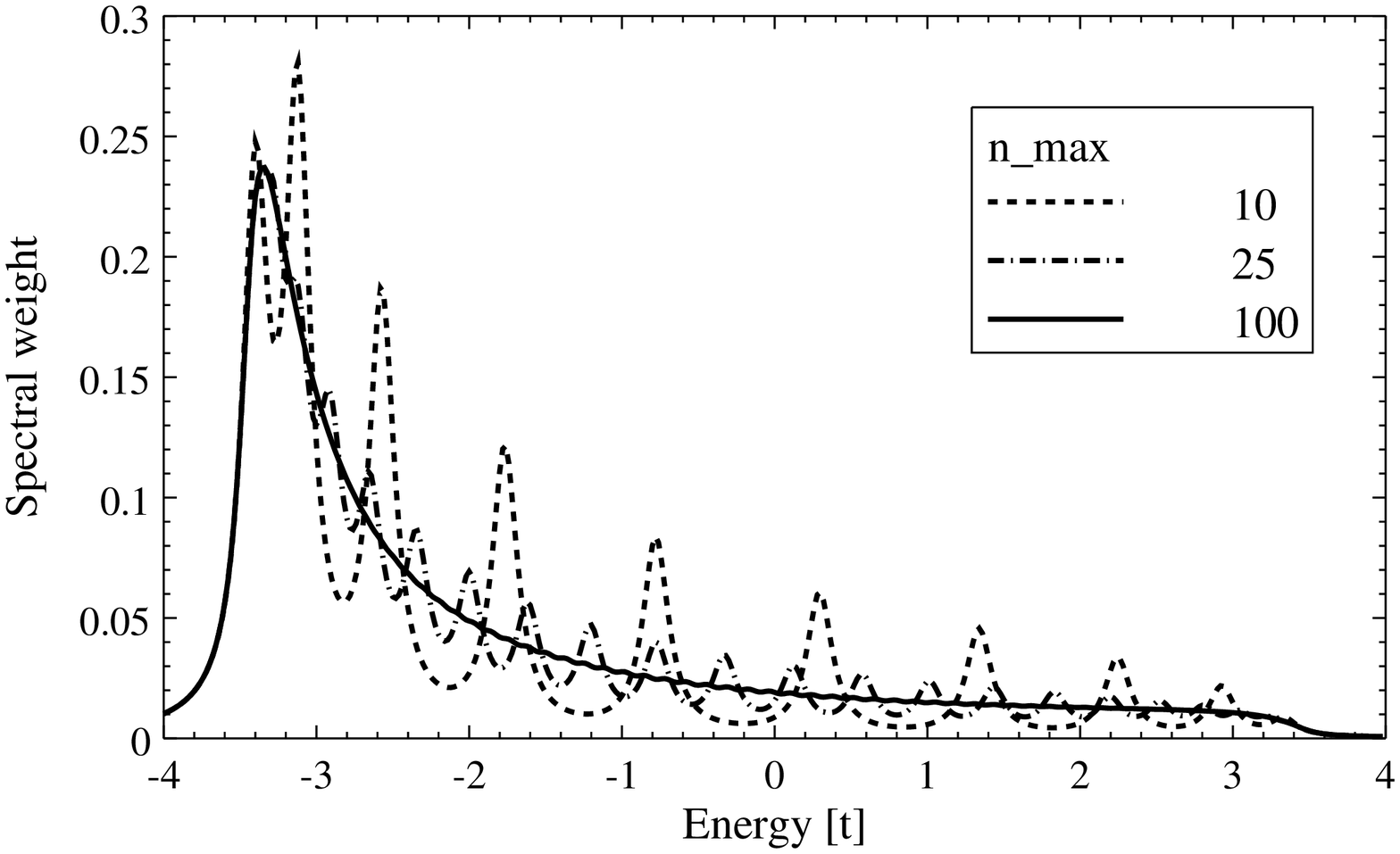}
\vspace{1cm}
\caption{ Effect of maximum path length $n_{max}$ on the spectral
          function. The curves are calculated for random background,
          $J=0$ and momentum $(\pi,\pi)$. Note that the artificial 
          linewidths are different for the three curves.}
\vspace{1cm}

\epsfxsize=13cm
\epsfysize=8cm
\epsffile{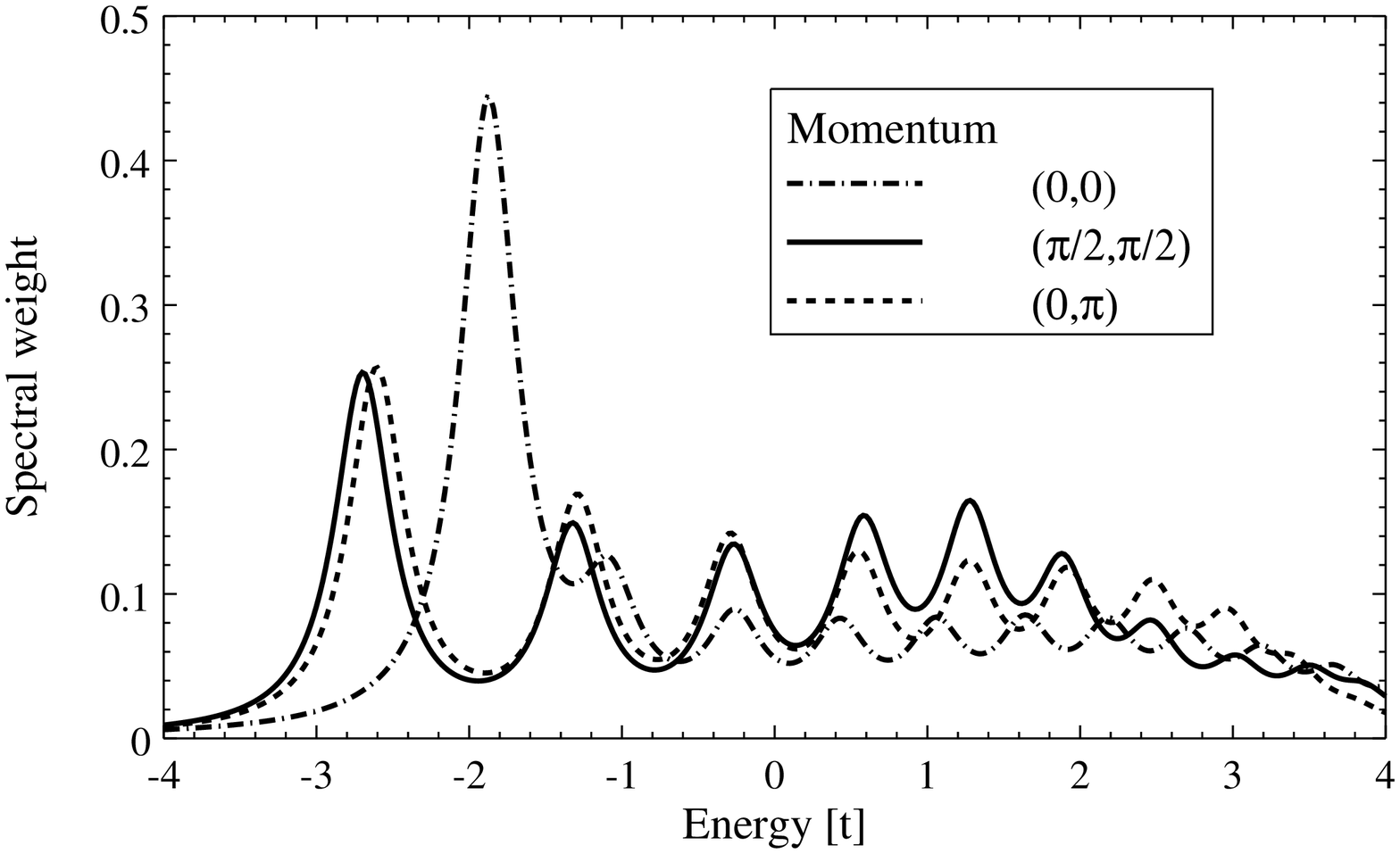}
\vspace{1cm}
\caption{ One-hole spectral function for the N\'{e}el-type background,
          $J/t = 0.4$ and different hole momenta. 
          We have used the reduced set (\protect\ref{REDVARSET})
          of dynamical variables with $n_{max}=256$ and a linewidth of
          $0.1 t$.
          The total bandwidth
          of the quasiparticle dispersion is $2J$ which is smaller than
          the correct value $2.2J$ which would be obtained with
          the full set of variables (\protect\ref{FULLVARSET}). }
\newpage

\epsfxsize=13cm
\epsfysize=8cm
\epsffile{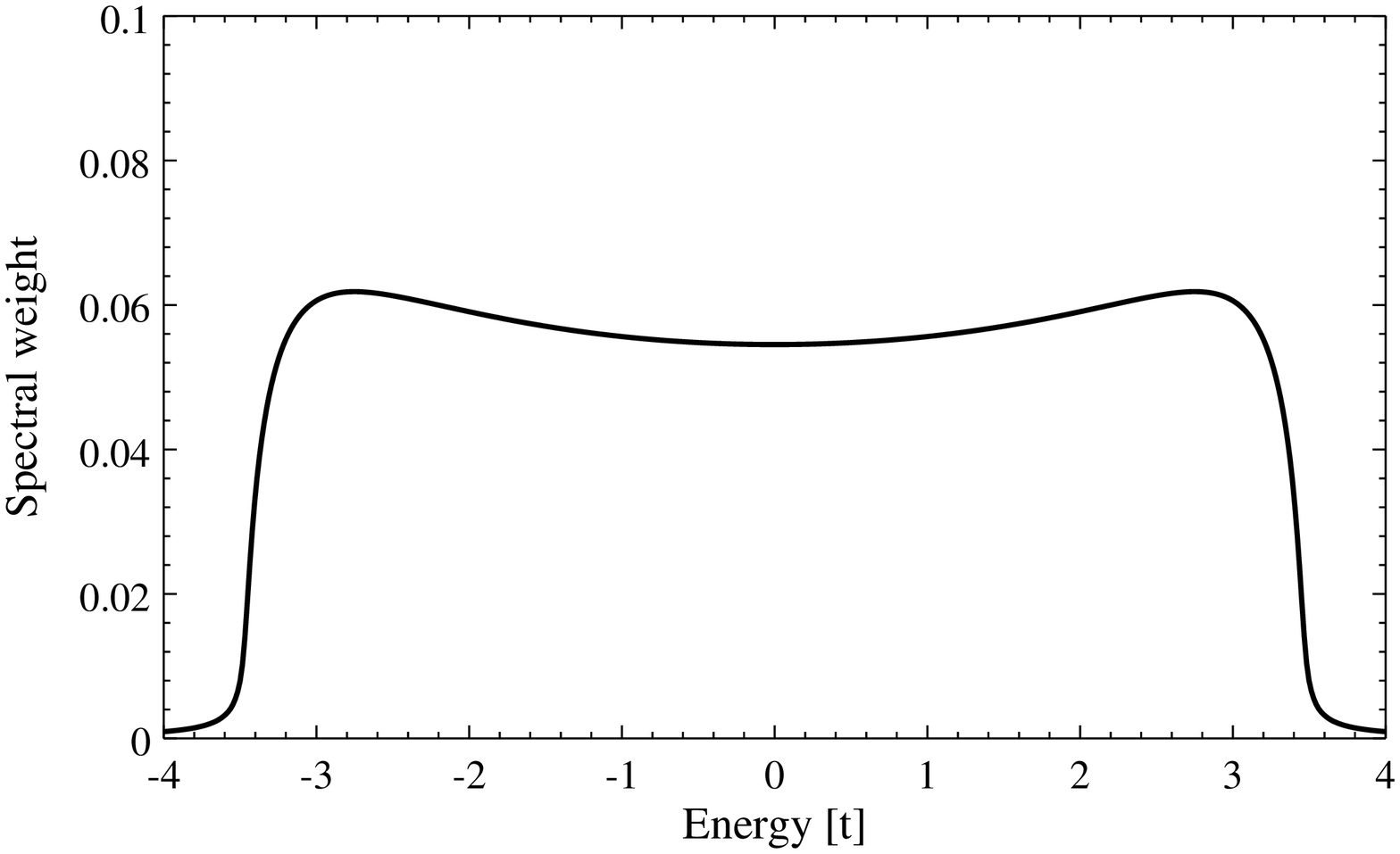}
\vspace{1cm}
\caption{ One-hole spectral function for the N\'{e}el-type background
          and $J = 0$. Note that this spectrum does not depend on the
          hole momentum since the hole is localized. }
\vspace{1cm}

\epsfxsize=13cm
\epsfysize=8cm
\epsffile{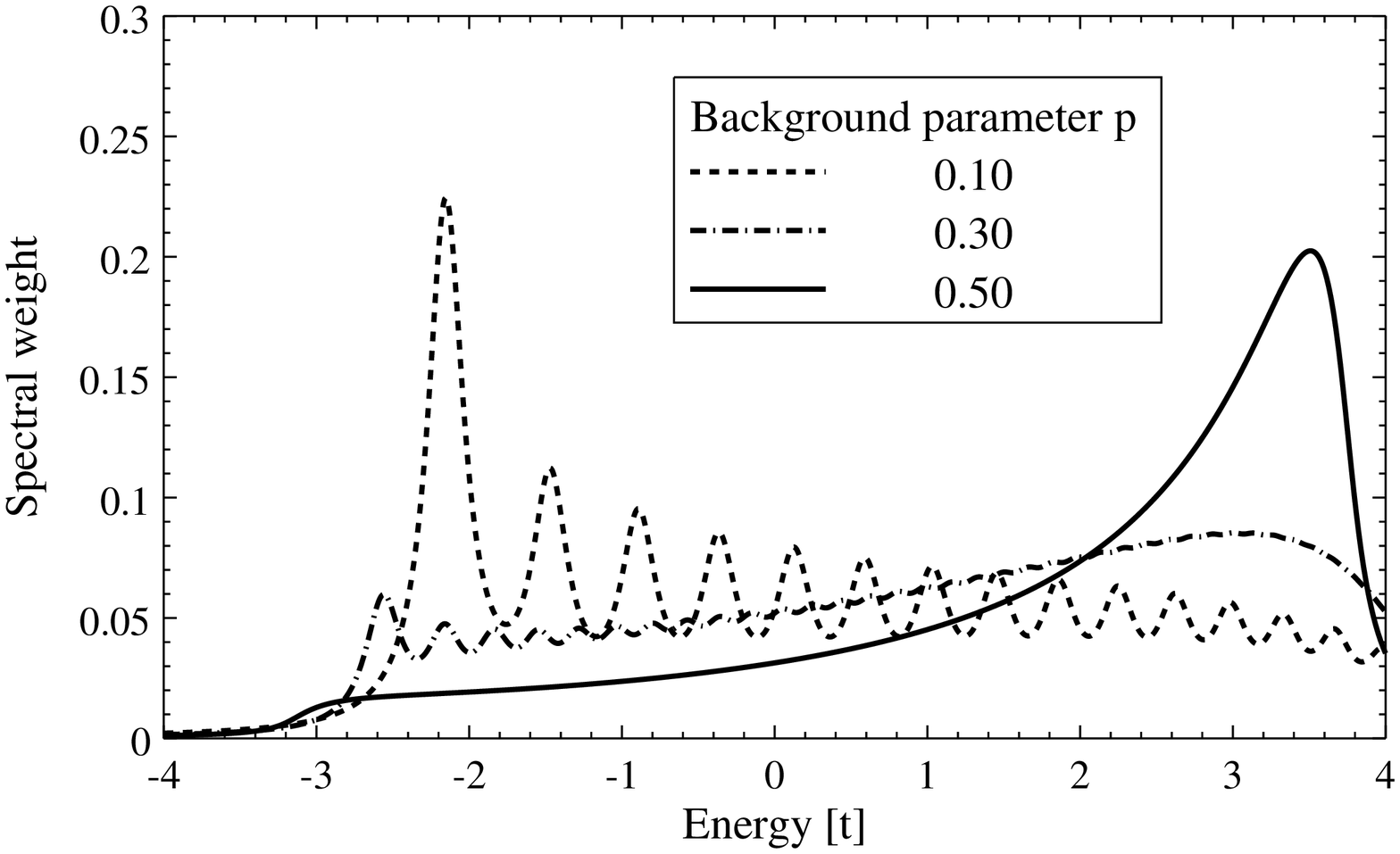}
\vspace{1cm}
\caption{ One-hole spectral function calculated using the interpolation
          ansatz (\protect\ref{INTERP_ANS}), $J/t=0.4$, the 
          mean-field term (\protect\ref{MEAN_J}), 
          momentum $(0,0)$ and different values of $p$. 
          The linewidth is $0.02 t$.}
\newpage

\epsfxsize=13cm
\epsfysize=8cm
\epsffile{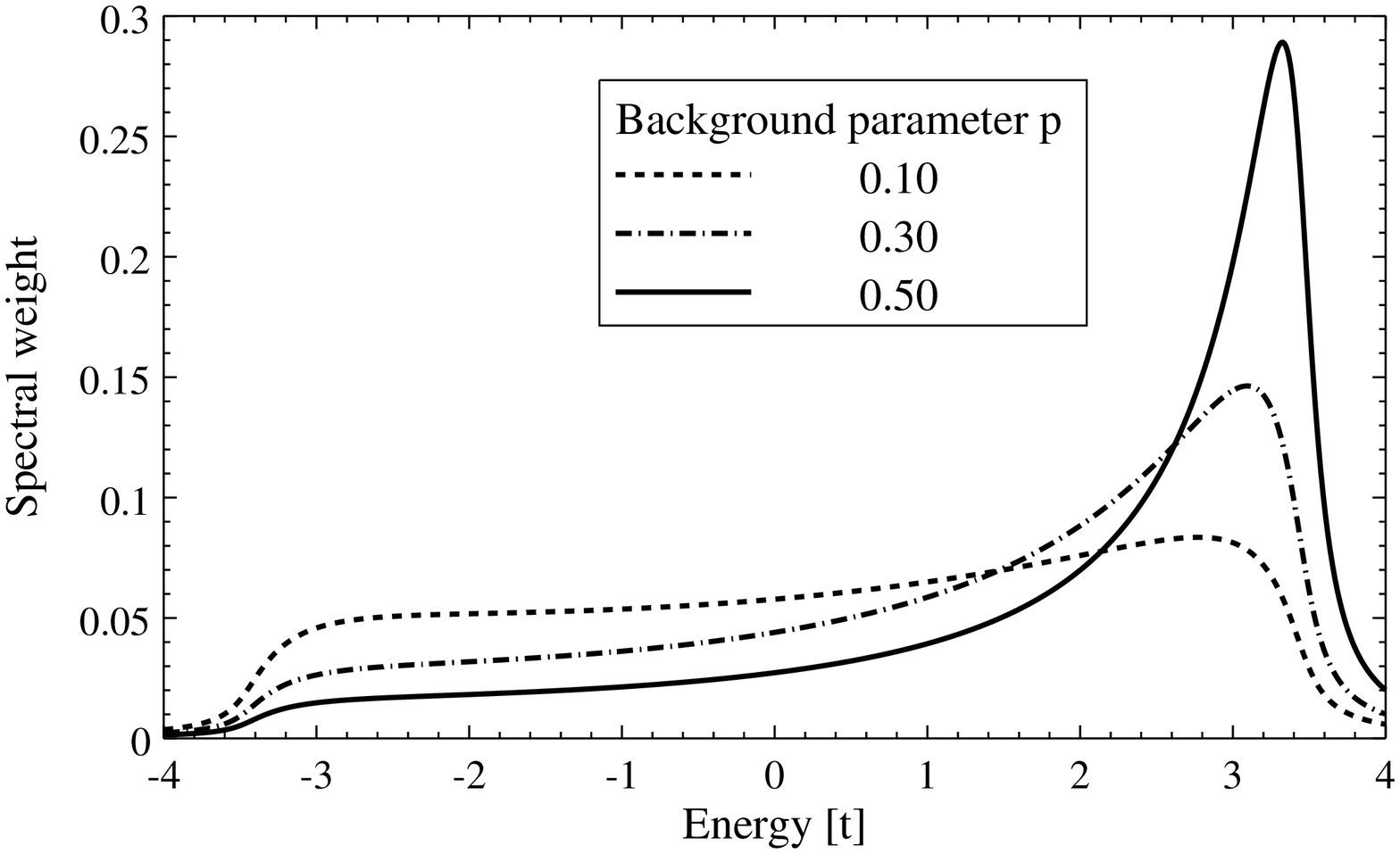}
\vspace{1cm}
\caption{ Same as Fig. 6, but for $J = 0$. }
\vspace{1cm}

\epsfxsize=11.5cm
\epsfysize=7.5cm
\epsffile{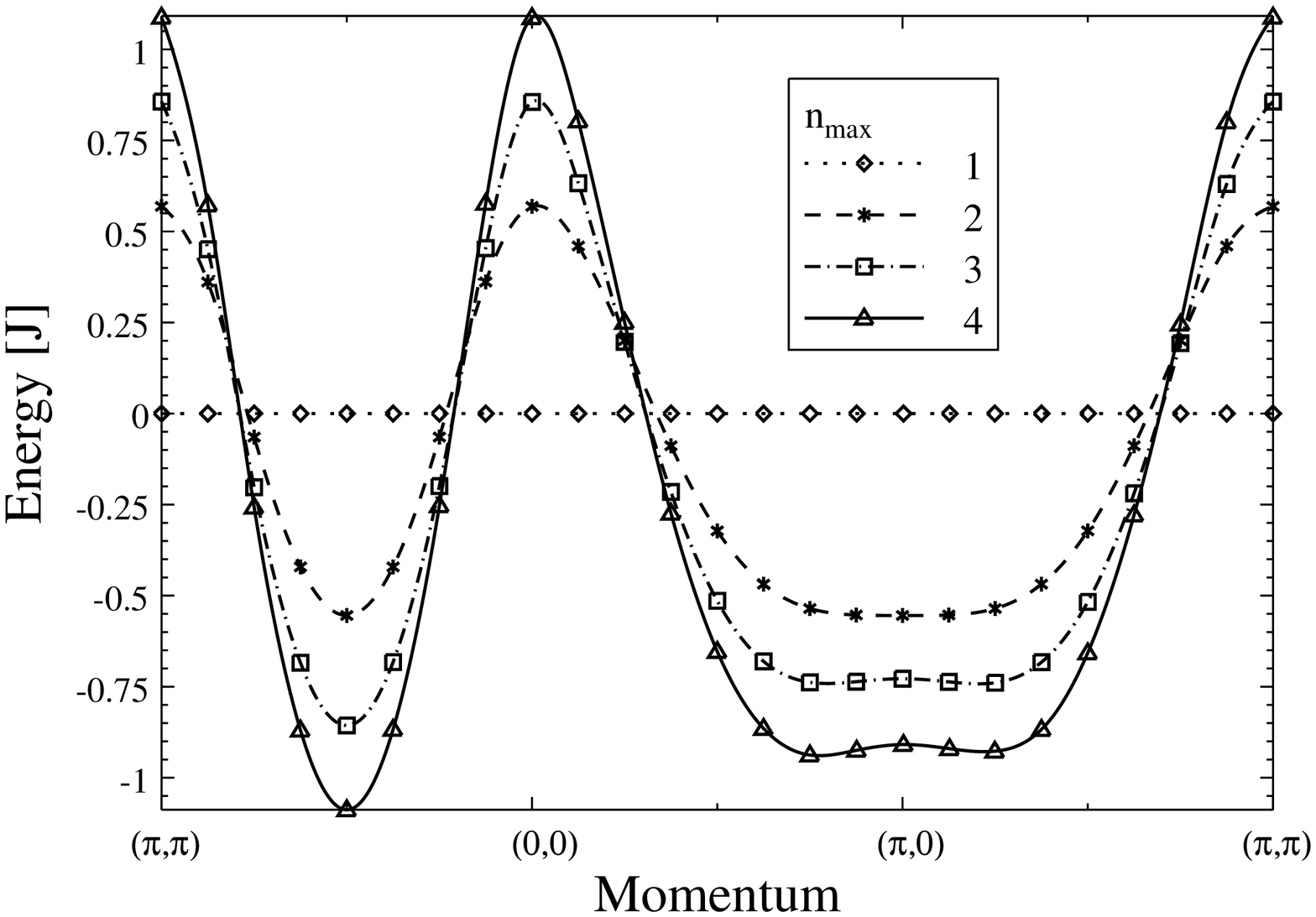}
\vspace{1cm}
\caption{ Dispersion of the lowest energy eigenvalue for $J/t=0.4$,  
          antiferromagnetic background and the full variable set 
          (\protect\ref{FULLVARSET}). Curves are shown for 
          maximum path lengths $n_{max} =$ 1,2,3, and 4. 
          The zero energy level has been set to the center of mass of 
          the band.}

\end{figure}

\end{document}